\documentclass[prb,preprint,showpacs,preprintnumbers,amsmath,amssymb]{revtex4-2}
\usepackage{graphicx}
\usepackage{dcolumn}
\usepackage{bm}
\usepackage[mathscr]{eucal}
\usepackage{mathrsfs}

\usepackage{txfonts}

\begin{document}

\title{Residual Entropy of Glasses and the Third Law Expression}


\author{Koun Shirai$^{1,2}$}

\affiliation{
$^1$ SANKEN, Osaka University, 8-1 Mihogaoka, Ibaraki, Osaka 567-0047, Japan \\
$^2$ Vietnam Japan University, Vietnam National University, Hanoi, \\
Luu Huu Phuoc Road, My Dinh 1 Ward, Nam Tu Liem District, Hanoi, Vietnam 
}

\begin{abstract}
The third law of thermodynamics dictates that the entropy of materials becomes zero as temperature ($T$) approaches zero. Contrarily, glass and other disordered materials exhibit nonzero entropy at $T=0$, which contradicts the third law. For over a century, it has been a common practice to evade this problem by regarding glass as nonequilibrium. However, this treatment causes many inconsistencies in thermodynamics theory. This paper resolves these inconsistencies and provides a rigorous expression of the third law without any exception. To seek the entropy origin, the anthropomorphic feature of entropy must be resolved. 
Because entropy can be uniquely determined only when thermodynamic coordinates (TCs) are specified, we have to identify these TCs. For this purpose, the equilibrium for solids should be clearly defined such that it does not depend on solid structures. On this basis, it is deduced that TCs of solids are the equilibrium positions of atoms. TCs comprise a thermodynamic state space, on which a unique value can be assigned to entropy. For solids, equilibrium states are specified by discrete points in the thermodynamic state space, which define atom configurations. Among various atom configurations, only one is thermally activated at sufficiently low temperatures, which is the equilibrium state at $T=0$. Other atom configurations are called frozen configurations, which do not contribute to the temperature dependence of entropy near $T=0$. The rigorous statement of the third law has been established by expressing that the entropy associated with the active configuration vanishes at $T=0$. Residual entropy arises when the entropy is evaluated on an extended state space including the frozen configurations, which were previously active at high temperatures. The reconciliation of the two different views is explained through several debates on the glass transition.
\end{abstract}

\maketitle

\section{Introduction}

\subsection{Historical background}
No other law in physics has provoked as much controversy, while still lacking a clear consensus, as than the third law of thermodynamics. The third law is commonly referred to as the Nernst heat theorem, which states that the entropy ($S$) of a material vanishes as temperature ($T$) approaches absolute zero. 
However, apparent exceptions to this law have long been recognized. Certain asymmetric diatomic molecular crystals, ice crystals, and solid solutions exhibit nonzero entropy at $T=0$. The finite entropy at $T=0$ is called residual entropy, $S_{\rm res}$. Early investigations of this phenomenon are well documented in standard textbooks \cite{Fowler-Guggenheim,Wilson,Lewis-Randall,Wilks,Guggenheim,Beattie}. 
A physical law that admits exceptions calls for careful examination. A consistent explanation is therefore required for the existence of exception. At early days, it was argued that the validity of the third law should be restricted only to chemically pure crystals. This restriction, however, appears to lack a solid foundation. For example, a silicon crystal is chemically pure, yet it possesses residual entropy arising from its isotope distribution. Conversely, many compound crystals exhibit vanishing entropy at $T=0$, despite not being chemically pure. Residual entropy associated with the mixing of different elements, which is observed universally in all random alloys, has posed a persistent challenge for the third law since its discovery \cite{Eastman33}.

Among these apparent counterexamples, the most important and thereby most widely studied material is glass; see a review \cite{Gutzow09}. Following extensive investigations in the early of 20th century, Simon concluded that glass is not in an equilibrium state below the glass transition temperature ($T_{g}$) and is therefore exemplified from the laws of thermodynamics \cite{Simon30,Simon51}. He justified this view by considering that thermodynamics laws are applicable only to equilibrium states \cite{comment1}. Since then, the nonequilibrium interpretation of glasses has become widely accepted \cite{comment2}. However, this explanation does not resolve all conceptual difficulties. Now, we know that residual entropy originates from freedom in atom configurations, expressed as $S_{0} = k_{\rm B} \ln W_{c}$, where $k_{\rm B}$ is Boltzmann's constant and $W_{c}$ is the number of configurations accessible to the system at $T=0$. This configurational freedom is associated with structural imperfection in general. Because no real crystal is completely free from defect, it follows that no crystals obey the third law. If the reasoning applied to glass were extended to defective crystals, one would be led to the untenable conclusion that all crystals are in nonequilibrium states. Moreover, with the development of material research, numerous materials possessing structurally ambiguous or nonperiodic order, such as incommensurate crystals, quasicrystal, and frustrate systems, have been discovered. Whether these materials should be regarded as equilibrium or nonequilibrium states remains an open question. 

The view that the current state of a glass is a nonequilibrium state is often rationalized by regarding that, given sufficient time, the glass will eventually crystallize or transform into an ``ideal" glass with vanishing residual entropy. However, there is no experimental evidence to support this expectation. Instead, substantial evidence indicates that glasses can remain essentially unchanged for periods exceeding millions of years \cite{Berthier16,Zhao13}, as discussed in Sec.~\ref{sec:equilibrium}.
A similar argument is invoked for random alloys, which are sometimes presumed to transform into ordered compounds over sufficiently long time. Yet this expectation also lacks direct experimental support. An extreme example is the random distribution of isotopes in a crystal: it is difficult to conceive that isotopes would spontaneously rearrange into an ordered configuration. Constructing a fundamental law on the basis of such speculative assumptions is not advisable. Furthermore, the implicit premise that the equilibrium state of a material must always be ordered is itself questionable. Frustrated systems, for example, possess no long-range ordered state. For quasicrystal and DNA, even the notion of ``order" is unclear and not uniquely defined \cite{Gell-Mann95,Gell-Mann96,SFworkshop89}. Since Schr\"{o}dinger described DNA as ``aperiodic crystal" \cite{Schrodinger44}---though the DNA structure was not discovered at that time, the true meaning of this self-contradictory word remains unresolved. In contemporary biological literature, it is common understanding that the entropy of DNA includes the configuration entropy \cite{Gatlin72,Brooks86,Wicken87}, which plays a central role in the theoretical treatment of biological evolution. The configuration entropy of DNA does not vanish at $T=0$, and thus the entropy of DNA violates the third law. This apparent inconsistency is rarely addressed explicitly.
Although one could devise distinct explanations for the residual entropy on a case-by-case basis, such explanations relegates the third law from the first-rank position of physics law. For this reason, some researchers questioned the general validity of the third law \cite{Landsberg56,Landsberg78,Landsberg97, Buchdahl,Hasse,Tolman79,Tisza,CAQ,Baierlein,Waldram} and even abandoned it from fundamental laws. 

Despite the above difficulties, there is a compelling reason to defend the third law. An alternative expression of the third law is the principle of the unattainability of absolute zero temperature. In this expression, the third law turns to be a universal law without any exception, irrespective of whether the system is in equilibrium. This unattainability cannot be deduced from other laws of thermodynamics, making the third law indispensable. Fowler and Guggenheim exploited the unattainability of absolute zero temperature to deduce the rigorous expression for the Nernst heat postulation that admits no exceptions \cite{Fowler-Guggenheim}. In a primitive sense, their third law expression is correct, if the frozen state is suitably interpreted \cite{Shirai18-res}. 
Several studies have examined whether the two expressions---the Nernst neat postulation and the unattainability of the absolute zero temperature---are equivalent \cite{Hatsopoulos, Hasse,Levine, Landsberg56, Landsberg78,Landsberg97,McNabb17,Masanes17}.
Hatsopoulos and Keenan pointed out that their equivalence cannot be established when the residual entropy is involved (\cite{Hatsopoulos}, p.~29 in Foreword). Without resolving the residual entropy problem, the proof for this equivalence remains incomplete \cite{Shirai18-Unattain}. 

Recently, debates concerning the residual entropy have been revived, in connection with more subtle problems of glass. Some computer simulations and thermodynamic considerations showed an uncompensated decrease in entropy at the glass transition, which would lead to zero entropy at $T=0$ \cite{Kivelson99,Speedy99,Mauro07}. These studies thus support the validity of the third law. However, this raises the question of why residual entropies are observed in experiments. Kivelson and Reiss attributed this apparent contradiction to the problem of calorimetric measurement. The entropy change $\Delta S=S_{B}-S_{A}$ from state $A$ to $B$ is obtained by the calorimetric method as
\begin{equation}
\Delta S = \int_{A}^{B} \left( \frac{dQ}{T} \right)_{\rm rev}
\label{eq:defS}
\end{equation}
where the subscript (rev) indicates that the integration must be performed along a reversible path. The glass transition is generally an irreversible process, and, according to these authors, the obtained calorimetric value does not reflect the true entropy \cite{Kivelson99}. This claim invoked serious debate \cite{Gupta07,Goldstein08,Reiss09,Goldstein11,Takada13,Moller06,Aji10, comment4}. 
A central issue in this debate is an uncompensated reduction of entropy. This reduction implies a spontaneous reduction of entropy, which contradicts the second law of thermodynamics. Gupta  and Mauro evaded this contentious problem in terms of a human intervention of observation: they argued that the glass transition occurs only because our observation time is finite, and that the act of observation serves a constraint that is not treated within the thermodynamic laws \cite{Gupta09}. 

Regarding the reality of residual entropy, the debate has largely converged on the conclusion that the traditional view---namely, the residual entropy is real---is correct. We are thus brought back to the original problem of contradiction between residual entropy and the third law statement. Moreover, throughout this debate, one has recognized the logical inconsistency in attributing the residual entropy of glass to the nonequilibrium character. If glass were genuinely in a nonequilibrium state, the very notion of the residual entropy itself would lose its thermodynamic meaning. The residual entropy is measured by the calorimetric method, and this manner presupposes equilibrium, as indicated by the condition in Eq.~(\ref{eq:defS}). No one ever solved this fundamental inconsistency. In this way, the third-law issue is fallen into a deep impasse.

\subsection{Contentions}
The problem of residual entropy has been reiterated for over a century without a definitive resolution. By deeply analyzing the failures of previous explanation of the third law (further details of the failures are described in the early version of the manuscript \cite{Shirai18-res}), the author came to find that the core difficulty arises from the ambiguous understanding of equilibrium. 
It is repeatedly asserted that glasses are in a nonequilbrium state. However, it is illogical to use the term nonequilibrium when there is ambiguity in definition of equilibrium. The usual argument for regarding glasses as nonequilibrium systems is that their current properties are not determined solely by $T$ and volume $V$ (or pressure $p$) but also depend on the past history of thermal and mechanical treatments. Yet this raises a fundamental question: why should $T$ and $V$ be regarded as the only independent state variables? (Here chemical reactions are excluded, and thus compositional variables are not considered.) Although such a description is adequate for gases, no general proof establishes its validity for solids. If the current state can be uniquely specified by adding new state variables, then the state should be regarded as an equilibrium in an extended thermodynamic space. This point was demonstrated for plastic deformations by Bridgman \cite{Bridgman50}.
Bridgman cautioned against hastily concluding that the laws of thermodynamics do not apply to our system, if at first we are unsuccessful in finding a set of parameters that determine an energy function (Ref.~\cite{Bridgman61}, p.~59). Instead, he advised to suspect that {\it we have not a complete list of parameters of state}.
This perspective makes it essential to determine which quantities should properly be regarded as state variables. However, this question immediately gives rise to a circular difficulty: state variables are defined only for equilibrium, yet equilibrium itself is characterized in terms of state variables. This dilemma has been noted by some authors (see, for example, Callen's textbook Ref.~\cite{Callen}, Sec.~1.5), but it has not been satisfactorily resolved.

Another issue related to the third law concerns the arbitrariness in entropy. There is arbitrariness in entropy \cite{Grad61,Grandy}, which is called the {\em anthropomorphic} feature of entropy \cite{Jaynes65,Wigner63} (first used by E. P. Wigner). 
A fundamental thermodynamic relation,
\begin{equation}
dS = \frac{1}{T}dU -\sum_{j=1}^{M} \frac{F_{j}}{T}dX_{j},
\label{eq:dS}
\end{equation}
specifies how the state variables $X_{j}$ are related to the entropy change ($U$: internal energy, $F_{j}$: the generalized force conjugate to $X_{j}$, $M$: the number of variables). This equation determines how $S$ changes by $X_{j}$ but does not fix the entropy origin. A paramagnetic salt is in an ordered state when the crystal structure is studied, and hence we conclude $S_{\rm res}=0$. However, when the magnetic properties are studied, a residual entropy appears due to random orientations of the spins. This argument introduces a seemingly subjective character to the physical quantity of entropy. Even eminent physicist E.~P.~Wigner remarked that entropy is not a property of a material, which motivated his use of the term anthropomorphism \cite{Jaynes65,Wigner63}. In information theory, the entropy of a system depends on our prior knowledge of that system; it is therefore a quantity that must be updated in light of new measurements \cite{Jaynes79,Wehrl78,SFworkshop89,Balian03,Caticha21,Cover-Thomas}. 
It is said that there are two schools teaching probability, objective and subjective probabilities (\cite{Callen}, p.~384); ever-ending arguments are continued \cite{Jauch03-MD2,Beauragard03-MD2,Denbigh81,Rosenkrantz83}. Now, we see many different ways of defining entropy \cite{Wehrl78,Balian03,NonextEntropy04}. Leaving aside these debates related to information theory, the arbitrariness associated with entropy must be eliminated at least from thermodynamics. Otherwise, the entropy origin cannot be uniquely determined.

The two issues discussed above are longstanding and well-known problems in thermodynamics, and they remain unresolved. The difficulties associated with third law cannot be overcome without directly addressing these fundamental questions.
Henceforth, we adopt the term {\it thermodynamic coordinates} (TCs) to denote state variables, since the term ``coordinate" is a suitable word for mathematical description of thermodynamic states \cite{Zemansky}.

\subsection{Orientation and organization of this paper}
By examining the problems of residual entropy, the present study aims at establishing an unambiguous statement of the third law. This paper constitutes a substantially revised version of the author's earlier works \cite{Shirai18-res, Shirai18-Unattain}. 
As noted above, the central issue underlying residual entropy is the consistent understanding of equilibrium and TCs, a goal that was not fully achieved in the previous studies. 
The rigorous definitions of equilibrium and TCs are given in author's another study \cite{Shirai18-StateVariable}, which is an extension of the theory of Gyftopoulos and Berreta \cite{Gyftopoulos} to solids. Here, the unambiguous expression of the third law is constructed by starting from the rigorous definitions of equilibrium and TCs. 

The rigorous definitions of equilibrium and TCs adopted here are so novel from the conventional understanding that readers may therefore question the validity of the present theory. 
Historically, many fundamental issues surrounding the third law have been discussed through the glass issues, and consequently a careful investigation of specific problems of glass is indispensable. In parallel with the study on the third law, the author has studied glass issues using the present theory as theoretical basis, and has obtained a number of consistent results. These include analyses of the glass state \cite{Shirai20-GlassState}, the hysteresis in glass transition \cite{Shirai21-GlassHysteresis}, the activation energy of the transition \cite{Shirai21-ActEnergy}, and the specific heat jump \cite{Shirai22-SH,Shirai22-Silica}. Schr\"{o}dinger's idea ``aperiodic crystal" for describing DNA can be also understood on this basis \cite{Shirai25-OrderParams}.
Although this paper is written to be as self-contained manner as possible, the physics of glass is so deep, the detailed analyses of the historical problems of glasses cannot be fully explained in this paper. For these details, refer to the above original papers. A recent review on the specific heat of liquids and glasses \cite{Shirai-SH-Liquids25} is particularly useful to understand how the present theory consistently accounts for the thermodynamic properties of these materials. 

One of the conclusions of Sec.~\ref{sec:Thermodynamic-space}---namely, that equilibrium positions of atoms are state variables for solids (Corollary 3)---may appear to deviate significantly from the traditional view that the strength of thermodynamics lies in its use of only a small number of macroscopic variables. Readers may therefore find this conclusion difficult to accept when reading just this conclusion. However, this statement reflects experimental facts: any change in the structure of a solid---for example, through the introduction of defects---necessarily leads to changes in its internal energy and specific heat. If an inconsistency arises between theoretical assumptions and experimental observations, the theory must be revised accordingly. The reader is therefore encouraged to follow carefully the logical development presented in this paper.
The traditional belief that thermodynamics should avoid use of microscopic quantities, such as atom positions, may further hinder acceptance of Corollary 3. However, no principle within thermodynamics---from the first law to third---explicitly inhibit the use of microscopic quantities. Indeed, the fundamental statements of thermodynamics and statistical mechanics are mutually consistent and, in appropriate sense, equivalent. See Supplemental materials (Q2 and Q3). Supplemental materials include further useful answers to questions that the author has received on various occasions.
With these preliminaries established, we now proceed to examine the residual entropy of glass.

This paper is organized as follows. In Sec.~\ref{sec:equilibrium}, a rigorous definition of equilibrium is presented. In Sec.~\ref{sec:Thermodynamic-space}, the definition of TCs is given, together with several relevant notions. In particular, the notion of frozen coordinates plays a crucial role in clarifying  the anthropomorphic feature of entropy. The essential elements of Secs.~\ref{sec:equilibrium} and \ref{sec:TC} were already given in a previous study \cite{Shirai18-StateVariable}. 
However, because the third law is of such fundamental importance, a logically complete exposition is indispensable. For this reason, these materials are included here, with some rearrangements. 
On the basis of this framework, an appropriate formulation of the third law is developed in Sec.~\ref{sec:third-law}. In Sec.~\ref{sec:glass-transition}, several intriguing arguments concerning residual entropy at the glass transition temperature are examined, thereby illustrating the advantage of this study. Finally, Sec.~\ref{sec:conclusion} summarizes the main results.

\section{Equilibria in a solid}
\label{sec:equilibrium}

\subsection{Definition of equilibrium}
Teachers of thermodynamics have difficulty in explaining equilibrium at the beginning of a course \cite{Callen}. Thermodynamic equilibrium is characterized by TCs, such as temperature and pressure. However, what exactly are TCs? One may answer that TCs are macroscopic properties characterizing equilibrium. But, the formers have sense only in the latter, rendering the argument circular \cite{Beauragard-note}.
The only way to escape from this circular argument is to start outside equilibrium and to treat equilibrium as a special case of nonequilibrium. This is the approach that Gyftopoulos and Beretta (GB) in establishng a consistent theory of thermodynamics \cite{Gyftopoulos}.
They placed the second law at the starting position of thermodynamics. Only the second law is more fundamental than such a fundamental notion of equilibrium, since a fundamental notion cannot be deduced from less fundamental laws. The GB approach is general and applies irrespective of the type of substance, whether gas or solid (a brief exposition of the GB approach is given in Supplemental material; however, for deeper understanding, refer to the original work \cite{Gyftopoulos}.)
Unfortunately, GB did not show how their theory can be applied to solids. This is by no means a trivial task. As noted in Introduction, every real crystal contains imperfections, which makes it difficult to uniquely specify equilibrium state. It is sometimes argued that, because defect states of a crystal are metastable, the principles of thermodynamics cannot be applied to them. However, it is our leaning from elemental thermodynamics that the equilibrium concentration of the defect is determined by the condition of the minimum in the free energy. In this sense, the perfect crystal is unstable nonequilibrium state. In fact, perfect crystals do not exist. 
Furthermore, there are a class of materials possessing ambiguous structures, such as frustrated systems. For such systems, even identifying the equilibrium state can be problematic. We are therefore compelled to reexamine the definition of equilibrium.

Let us begin to consider a system consisting of many particles. In general, the system is in a nonequilibrium. Such a state is characterized by a set of time-varying (dynamical) variables $X(t)$: any physical quantity determined by well-defined measurements qualifies as a dynamical variable. The number of particles in a small segment of space, $n({\mathbf r},t)$, is an example. 
Suppose that the system is adiabatically connected to a weight as the sole device external to the system.

\noindent
{\bf Definition 1: Thermodynamic equilibrium}

{\em It is impossible to change the stable equilibrium state of a system to any other state with its sole effect on the environment being a raise of the weight}. [GB, p.~58]

\noindent
No reference to TCs appears in this definition; hence, no circularity arises.
The expression ``with its sole effect on the environment being a raise of the weight" means in usual terms that work can be extracted from the system without leaving any other effect on the environment. Definition 1 may be regarded as a generalization of Clausius's statement of the second law, which asserts that it is impossible to extract work from a single heat bath. A single heat bath is replaced by an equilibrium state in this definition. 
If work were extracted from such a bath, the energy of this work must come at the expense of a decreasing internal energy of the bath, since no other source is available. The decrease in the internal energy means cooling of the bath. It turns out that work would have been extracted solely by cooling a system. This contradict with Clausius's statement.
If two systems $A$ and $B$ are in mutual equilibrium, bringing them into contact does not cause any change in their states. If they are not in equilibrium, the contact induces a spontaneous change accompanied by entropy production. The maximum work theorem (\cite{Callen}, \S 4.5) states that an increase in entropy can be converted into net work delivered to the environment, if an appropriate thermal engine is inserted. If a system has hot and cold regions, heat flows from the hot to the cold region, and net work can be extracted from this flow. Once a homogeneous temperature distribution is established, no further work can be obtained. The equality of temperature between two regions in mutual equilibrium agrees with the zeroth law of thermodynamics.

From Definition 1, it follows that all the existing crystals are in equilibrium states, since no work can be extracted from a crystal without affecting the environment. Glass is no different from crystals in this respect. If work were obtained from a glass without affecting the environment, this would violate the second law. The equilibrium characterization of glass thus follows directly from Definition 1. 
In material science, glass is generally regarded as being in nonequilibrium state. One reason that they consider is the sluggish structural evolution of glass. This issue is discussed in the next subsection in connection with timescale. Another reason is that the current properties of glass cannot be expressed solely by the current value of $T$ and $V$, as noted in Introduction. However, the assumption that $T$ and $V$ are the only independent TCs for solids is unfounded. This issue constitutes the main topic of Sec.~\ref{sec:Thermodynamic-space}. Furthermore, it is often claimed that the glass state is less stable than the supercooled liquid at $T<T_{g}$. This assertion is not supported by thermodynamic principles: the free energy of a glass is lower than that of the supercooled liquid state \cite{Shirai20-GlassState,Shirai-SH-Liquids25}. Within the framework of thermodynamics, Definition 1 alone represents the equilibrium concept, which is independent of material's internal structures.

\subsection{Constraint, timescale, and ergodicity}
\label{sec:constraint}
\paragraph{Constraint}
In Definition 1, the phrase ``its sole effect on the environment" signifies that the only external effect is the raising of a weight. This implies that all constraints are fixed. Consider, for example, a gas confined in a cylinder with a piston. When the gas is homogeneously distributed, it is in equilibrium, provided that the piston is held fix. However, if this constraint is relaxed, work can be obtained. Fixing constraints is thus essential in the context of Definition 1. In the above case of the gas, the constraints are clearly identified as the walls of the container, which specify the volume of gas. However, inside materials, the notion of constraints becomes is not always equally transparent.

The idea of constraints in materials was introduced by Gibbs more than a century ago \cite{Gibbs-TD}, who used the term passive resistance. His intention was to distinguish two categories in static states.
One case is a static state that is achieved by the balance of the active tendencies of the system. There is another case of static states in which a static state persists because any change is prevented by passive resistances. Paraffin oil is stable at standard temperature and pressure, and the paraffin oil and air coexist in equilibrium. A constraint that inhibits the chemical reaction is operative in this case.
The nature of constraints was not known in Gibbs's time; today, they are recognized as an energy barrier. An energy barrier suppresses the reaction rate to such an extent that the reaction is effectively inhibited. A passive resistance is a perfect inhibitor, contrary to its modest name \cite{Hatsopoulos}. In the present context, however, there is no need to distinguish the above two categories of equilibria. Both macroscopic walls and microscopic energy barriers in materials function as constraints.

\noindent
{\bf Definition 2: Constraint}

{\em A constraint $\xi_{j}$ is a means of restricting the range in which a particular variable $X(t)$ can vary. }

\noindent
It is expressed either by a hypersurface $\xi_{j}(X(t))=0$ or by a range $\xi_{j}(X(t))>0$. 
Although such a restriction can be applied to any mode of change, we restrict, in this paper, ourselves to those in the real space. 
For a solid, atoms cannot move beyond the unit cell to which they belongs. The boundary, $\xi_{j}$, of the $j$-th unit cell for the $j$-th atom plays the role of a constraint. However, a more realistic substance of the constraint in this case is the energy barrier $E_{b,j}$ that surrounds the $j$-th atom. 
A crystal retains its structure through a set of constraints to which every atom in the crystal is subjected to. This is also true for nonperiodic materials. In this respect, the structure is akin to a constraint, as it distinguishes one type of solid from others. In the glass literature, a special term {\it kinetic constraints} is often used \cite{Reiss,Kivelson99,Corti97}. This term is regarded as an artificial device in order to make thermodynamic analysis amenable. In this study, constraints are real substances that refer to energy barrier inside materials.

\paragraph{Timescale}
Any equilibrium state is defined with respect to a particular timescale. 
The common view for glass is that it undergoes internal structural changes, namely atomic rearrangements, but at an extremely slow rate. There is even an extreme view that the glass is a kind of liquid with infinitesimally small flow velocities \cite{Zanotto17}. However, it is important not to overlook that no equilibrium is eternal. For example, there is no doubt to regard a gas in a container as being in equilibrium, if it is uniformly distributed. In reality, however, even a gas cylinder made of robust metal cannot retain its high-pressure state indefinitely. The gas eventually leaks out. Similarly, a glacier appears static on the timescale of a human lifetime, yet it flows on geological timescale. Thus, every equilibrium state has a finite life time; equilibrium is always defined relative to the observation timescale.

An equilibrium state is sustained by the given constraint $\xi_{j}$, which inhibits the change of the $j$-th type in the structure. The physical substance of a constraint is the energy barrier $E_{b,j}$ for the $j$-th type change. Real barrier heights are always finite, regardless of how large they are. Consequently, there exists a finite relaxation time, $\tau_{j}$, over which the constraint is effectively operative. The characteristic timescale $\tau_{j}$ of the equilibrium is governed by the energy barrier $E_{b,j}$ according to 
\begin{equation}
1/\tau_{j} = \nu_{j} e^{-E_{b,j}/k_{\rm B}T},
\label{eq:tau}
\end{equation}
where $\nu_{j}$ is an appropriate attempt frequency of transitions. 
Most metals degrade within a few hundred years due to corrosion, oxidation, and other effects. 
In contrast, glass is remarkably stable over very long timescales. The current observation indicates that volcanic glasses can preserve their structures for more than a million years, and the Moon beads have existed for more than a billion years (see Column in \cite{Berthier16}). This extraordinary stability is the reason why glass is used as a capsule material for nuclear wastes containment. 
The equilibrium character of glass is further supported by studies of its individual properties \cite{Shirai20-GlassState,Shirai21-GlassHysteresis,Shirai22-SH}.
It is inappropriate to use logically inconsistent arguments for constructing fundamental laws, and accordingly we do not use the nonequilibrium assumption for glass in the following.

\paragraph{Ergodicity-breaking process}
Equilibrium means ergodicity. An ergodic state is understood as one in which every particle in a system visits everywhere in the system with equal probability. However, this equal probability is often misconstrued, when constraints break equal probabilities. Penrose noted that, for nonergodic systems, the energy manifold can be partitioned into invariant submanifolds (\cite{Penrose79}, p.~1947). 
It is often stated that the glass transition is an ergodicity-breaking process. However, every crystallization process is also an ergodicity-breaking process: every atom becomes constrained such that it cannot move beyond the unit cell to which it belongs. In this sense, crystals represent the most ergodicity-breaking systems, and yet they are unquestionably in equilibrium.

Notably, even an ideal gas confined in a container breaks ergodicity in the above sense; the molecules inside the container cannot visit the space outside it. Of course, we understand that this restriction arises from the wall of the container, which prevent the molecules from escaping. However, remember that, in a real container, a gas leak always exists. Some molecules have visited the outside the container. Even if the wall were removed, another constraint would still remain. Air molecules cannot escape beyond the atmosphere of the earth. Their spatial distribution continuously changes under the influence of the gravitational potential $V(\bf r)$. Perfect ergodicity does not exist. Ergodicity, like equilibrium, is always defined relative to the constraints imposed on the system.

\subsection{Metastable, stable, and frozen states}
In material science, it is often useful to distinguish metastable from stable states. The state with lowest energy is ordinarily referred to as the stable state. However, in thermodynamics, the distinction is meaningful only in a relative sense.
Diamond is the most stable material on our planet. However, at room temperature, it is metastable compared with graphite: because the free energy of diamond is slightly higher than that of graphite \cite{Bundy96}. A mixture of hydrogen and nitrogen gases appears very stable, as no reaction occurs within a human lifetime at room temperature. Yet on sufficiently long timescales---such as those in the revolution of the sun system, the formation of ammonia is thermodynamically favored, as indicated by the difference in free energy. If nuclear reactions are taken into account, nearly all nuclei except iron are metastable. In this sense, there is no fundamental difference between stable and metastable equilibria; the difference lies only in the height of the relevant energy barriers and the associated relaxation times.
A metastable state $j$ becomes the stable state within the scope of a given problem, if the observational timescale is shorter than the relaxation time $\tau_{j}$.

Normally, defect states of a crystal are regarded as metastable states. The ground-state energy of a defective crystal is higher than that of the perfect crystal (here we restrict our discussion to intrinsic defects, such as vacancy and interstitial, without foreign atoms). However, once formed, the defect structure remains unchanged unless it is annealed out at sufficiently high temperatures. Consequently, within the relevant timescale, such a state must be regarded as an equilibrium state. 

\noindent
{\bf Corollary 1: Equilibrium states of a solid}

{\em If a solid does not change the current structure during a time period $\tau$, this state is an equilibrium state within the time period $\tau$, regardless of whether the structure contains defects.}

\noindent
By creating defects, numerous microscopic structures become possible for a given crystal. Accordingly, a crystal possesses as many equilibrium states as the numbers of atom arrangements that can persist over the timescale of interest.
By virtue of Corollary 1, we need not ask whether quasicrystals are in equilibrium states; if their structures remain unchanged over the relevant timescale, they qualify as equilibrium states in that sense.. 

\begin{figure}[ht!]
\centering
\includegraphics[width=.35 \textwidth, bb=0 0 339 241]{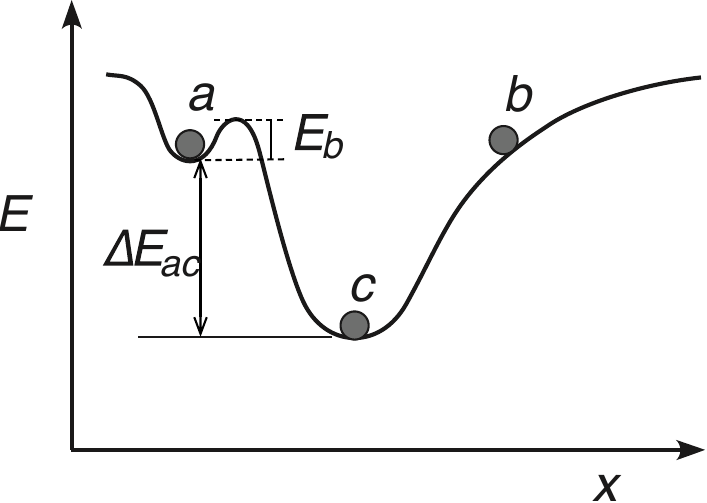}
\caption{
Distinction between metastable stable (a) and frozen state (b). (c) is the stable state.
}
\label{fig:meta-frozen} 
\end{figure}

This consideration applies particularly to glasses. We need not investigate which structure is the equilibrium state of a glass. Any glass currently existing is in an equilibrium state until its structure changes.
A few remarks are in order regarding glasses. It is often asserted that the glass state is neither a stable nor metastable state. Instead, the special term {\it frozen} state is used to differentiate from the aforementioned metastable state.
It is explained that the frozen state is essentially a nonequilibrium state that is immobilized by a strong viscous resistance. This conceptual distinction may be best illustrated in Fig.~\ref{fig:meta-frozen}: it is taken from Wilks's textbook (\cite{Wilks}, p.~61). In this figure, a frozen state appears to lack an energy barrier; instead, atomic motion is impeded by classical resistance. However, upon closely examination, the microscopic origin of viscous resistance is nothing other than an energy barrier: frictional forces do not exist in the microscopic scale (see Ref.~\cite{Shirai-SH-Liquids25}, \S 3.3.3). 
Furthermore, the term {\it kinetically frozen state} is also frequently used to emphasize the special property that the glass state is obtained by kinetic arrest from a supercooled liquid. However, the kinetic energy and potential energy are simply components of the total energy, and they are continuously interchanged during the course of microscopic dynamics. 
The substance that freezes atom motions is the potential barrier, which is built in by the reduction of kinetic energy as the temperature decreases. Any kind of crystallization is a freezing process of the atoms of a liquid. The difference between glasses and crystals lies in the quantitative aspect that a glass has an overwhelming number of metastable states whose energies are close each other. Now, glasses can be obtained also by vapor deposition methods \cite{Berthier16}. In this case, no supercooled liquid is involved. The expression {\em kinetically frozen liquid} therefore loses its literal meaning.

From Eq.~(\ref{eq:tau}), we see that the equilibrium timescale has strongly temperature dependence. As $T$ approaches 0, any value of energy barrier becomes infinitely large relative to the thermal energy $k_{\rm B}T$. Hence, we conclude that

\noindent
{\bf Corollary 2: Equilibrium at $T=0$}

{\em As the temperature approaches zero, all existing solids retain their equilibrium states for an effectively infinite time, unless the environment is altered. }

\noindent
At $T= 0$, the distinction among stable, metastable, and frozen states disappears. When discussing the third law, it is often argued that metastable or nonequilibrium states cannot reach the stable equilibrium state within a reasonable timescale at very low temperatures. In the present context, however, such states are already considered stable equilibrium states at low temperatures, because no structural change occurs over any finite timescale. At high temperatures, by contrast, transitions between different structures can occur within a relevant timescale. In that regime, the distinction between metastable and stable states becomes meaningful, because the timescale comes into our scope.

\section{Thermodynamic state space}
\label{sec:Thermodynamic-space}

\subsection{Thermodynamic coordinates}
\label{sec:TC}

\subsubsection{General considerations} 
\label{sec:TCgeneral}

By defining equilibrium by Definition 1, state variables can be introduced without falling into circular argument. Even when a system reaches an equilibrium state, its dynamical variables have time dependence due to thermal fluctuations. The details of such fluctuations are irrelevant with the thermodynamic properties of a system. The word {\it thermodynamic relevance} becomes important, as seen below.
These thermal fluctuations can be eliminated by taking a time average. Hence, it is good to define TC by time average of a dynamical variable $X(t)$, as suggested by Callen (\cite{Callen}, p.~5).

\noindent
{\bf Definition 3: Thermodynamic coordinate}

{\em A thermodynamic coordinate is the time-averaged value of a dynamical variable under a constraint that restricts the range within which the instantaneous value of the dynamical variable can vary}. 

\noindent
In a mathematical form, a TC, $X_{j}$, is written as
\begin{equation}
X_{j} = \frac{1}{t_{0}} \int_{\xi_{j}} X(t) dt,
\label{eq:t-aveX}
\end{equation}
where $t_{0}$ is the time interval over which the average is taken. $t_{0}$ must be shorter than the relaxation time $\tau_{j}$; otherwise, it can be chosen arbitrarily.
The notation $\xi_{j}$ in the integral indicates that the time evolution of the dynamical variable $X(t)$ is restricted by the constraint. By defining a TC in this manner, it is understood that the time average of the corresponding variable $X(t)$ converges to a finite value $\bar{X_{j}}$ and is independent of the specific choice of $t_{0}$, provided that $t_{0}<\tau_{j}$. This property of a TC is referred to as {\it definiteness}. 
For gas systems, the time average of the molecule's position, $\overline{x(t)}$, is indeterminate. Thus, $\overline{x}$ cannot be a TC for gases. In a more restrictive sense, there is a special case in which the molecules of a gas have definite averaged value $\bar{x}$. In a fixed container, the center of mass of a gas system gives the averaged position $\bar{x}$, when the gas molecules are uniformly distributed. However, the center of mass is irrelevant with the thermodynamic properties of the gas, and, for now, it is safe to exclude this trivial coordinate from a set of TCs for this reason of thermodynamic relevance. However, the problem of thermodynamic relevance is discussed deeply in Sec.~\ref{sec:Frozen-coordinates}.
An obvious feature of TC is that the time-averaged value $\bar{x}$ does not change whenever measurement is performed, as long as an equilibrium state continues. This constancy is here called the {\it sampling invariance}. This sampling invariance has a leading role in identifying {\it order} in materials \cite{Shirai25-OrderParams}.

Next consider the spatial distribution function, $\delta(x', x(t))$, of a molecule at $x'$ of a gas system. The integration of this function over the interior of the box,
\begin{equation}
X(t) = \int_{V} \delta(x', x(t)) dx',
\label{eq:int-distribution}
\end{equation}
yields the density-weighted volume. In equilibrium, the spatial distribution is uniform and thus the time-averaged value $\overline{X(t)}$ equals the volume $V$. The time-averaged value $\overline{X(t)}$ thus quantifies as a TC, namely, $V$.

A constraint determines a TC. For example, the constraint of a rigid wall $\xi_{1}$ of a cylinder specifies the volume $V$, which serves as a TC  for the gas inside the cylinder when the gas is in equilibrium. 
If an additional wall $\xi_{2}$ is inserted inside the cylinder, the system is divided into two regions, and two volumes, $V_{1}$ and $V_{2}$, arise as independent TCs. Similarly, inserting a further wall $\xi_{n}$ creates a new TC, $V_{n}$. In this way, we see that there is a one-to-one correspondence between constraints and TCs. This relationship between a TC and a constraint is well described by Reiss (\cite{Reiss}, Sec.~I.8). 
This example illustrates that there is no fundamental restriction on the number of TCs.
In the traditional formulation of thermodynamics, the thermodynamic properties of a macroscopic system are described by only a few TCs (see, for example, \cite{Callen}, p.~6). However, we must be more careful when enumerating TCs for solids. It is unclear how small the ``small" is. An infinite is a large number, yet the number of countable numbers is far smaller than the number of real numbers. In the statements of the thermodynamics laws---from the zeroth to the third---there is no principle that limits the number of TCs. Rather than such an ambiguous word, the present argument shows that existence of the one-to-one correspondence between constraints and TCs is more important \cite{comment-few-GB}.

Although any variable to meet Definition 3 can serve as a TC, imposing an additional requirement is desirable to maintain consistency with the normal usage of state variable. This additional requirement for TC is {\it distinguishability} from other TCs. To illustrate this point, consider again the above example of the gas in a container. Let us define, for an individual particle $k$, a dynamical variable $X_{k}(t) = \int_{V} \delta({\bf x}'-{\bf x}_{k}(t)) d{\bf x}' $. At equilibrium, the time average $\overline{X_{k}(t)}$ yields the same value $V$ for every particle $k$. There is no distinguishability among all the particles. The coordinates $X_{k}$ are redundancy. Therefore, a single TC---namely, the volume $V$---is sufficient to specify the thermodynamic properties of the gas.

\subsubsection{Thermodynamic coordinates of solids}
\label{sec:TCsolid}
For a gas, $V$ is only a TC for a fixed $T$. Any change in the shape of a container does not affect the properties of the gas, as far as $V$ is kept constant. In contrast, for solids, any deformation of shape alters the internal energy $U$, even if the volume is unchanged. Such a deformed solid is nevertheless in equilibrium, provided that the deformation is sustained by external constraints, such as nano-indenters (see Fig.~1 of \cite{Shirai18-StateVariable}). Another example is plastic deformation of a metal. In this case, the metal does not recover the original shape, even after the applied stress is removed. Microscopically, atom rearrangements involving dislocations occur \cite{Kestin70,Berdichevsky06}. The metal establishes a new equilibrium state with a different atomic configuration. 
Even a displacement of a single atom in a solid---for example, by electron irradiation---changes its $U$. 
In all these cases, once the solid reaches a new equilibrium state, each atom occupies a new equilibrium position $\bar{\bf R}_{j}'$, regardless of whether the position corresponds to a regular position.

The equilibrium position $\bar{\bf R}_{j}$ of an atom in a solid is the time average of its instantaneous position, ${\bf R}_{j}(t) = \bar{\bf R}_{j} + {\bf u}_{j}(t)$, where ${\bf u}_{j}(t)$ is a small displacement from the equilibrium position. Each atom thus possesses a definite value $\bar{\bf R}_{j}$. Moreover, the equilibrium position $\bar{\bf R}_{j}$ is distinguishable from those of all other atoms. Therefore, $\bar{\bf R}_{j}$ meets the requirements of definiteness and distinguishability established for TCs. 

\noindent
{\bf Corollary 3: Thermodynamic coordinates of a solid} 

{\em An equilibrium position of each atom $\bar{\bf R}_{j}$ is a thermodynamic coordinate for a solid.}

\noindent
Although the number of atoms in a solid is very large, it remains vastly smaller than the number of microscopic states. Even for a single atom $j$, the set of instantaneous positions ${\bf R}_{j}(t)$ constitutes infinitely many microstates, while the time-averaged position ${\bf R}_{j}$ is a single coordinate. Hence, the present conclusion is still consistent with the spirit of thermodynamics that a state variable characterizes thermodynamic properties of a system that has a huge number of microscopic states.

In addition, chemical species $\kappa$ at each position can be also a TC. There is no difficulty in treating this variable \cite{Shirai18-StateVariable}. However, in the following discussion, we neglect this variable to avoid unnecessary complexity, and restrict our attention to single-component solids.

\subsubsection{Fundamental relation of equilibrium}

In terms of the equilibrium state, the second law is expressed as follows:

\noindent
{\bf Postulation 1: The second law of thermodynamics}

{\em Among all the states of a system that have a given $U$ and are compatible with the given constraints, there exists one and only one stable equilibrium state.}  [GB, p.~63]

\noindent
In terms of entropy, the stable equilibrium state is the state of the maximum entropy. Under a given set of constraints, the system evolves spontaneously toward this maximum-entropy state. This spontaneous process automatically locates the equilibrium state uniquely. This view becomes particularly significant in investigating equilibrium states of a solid. Now, in solid state physics, it is widely known that a solid has a large number of local minima in the potential. Such a metastable state as (a) in Fig.~\ref{fig:meta-frozen} occurs ubiquitously; this is why the structural search is currently an active field \cite{Oganov06,Wang14}. In molecular dynamic simulations, we know that, for any initial atom configuration like (b) in the figure, an adiabatic relaxation process leads to one and only one local energy-minimum state. The final state, $(a)$ or $(c)$, depends on the initial state (b). Both states $(a)$ or $(c)$ are equilibrium states. Various glass states are obtained by different thermal history. All these final states are equilibrium states, if the final state meets the condition of Corollary 1. In this way, we can understand in the thermodynamic context that a glass has an almost infinite number of equilibrium states.

Postulation 1 states that there is a one-to-one correspondence between the set of constraints $\{ \xi_{j} \}$ and the resulting equilibrium state. In Sec.~\ref{sec:TCgeneral}, we established that there is a one-to-one correspondence between $\{ \xi_{j} \}$ and $\{ X_{j} \}$. It follows that the equilibrium state of a system is completely and uniquely specified by the set $\{ X_{j} \}$. Conversely, an equilibrium state fixes only one value $X_{j}$ for each $\xi_{j}$. 

\noindent
{\bf Corollary 4: Existence of the fundamental relation of equilibrium}

{\em Under given constraints with a fixed $U$, the thermodynamic coordinates $\{ X_{j} \}$ of a system are uniquely determined when the system is in equilibrium.}

\noindent
The statement in Postulation 1 that there exists one and only one equilibrium state for a given set of constraints implies the existence of a state function that characterizes that equilibrium. This state function is entropy $S$. Thus, there exists a function that relates $S$ uniquely to the complete set of TCs,
\begin{equation}
S=S(U, X_{1}, \dots, X_{M}).
\label{eq:FundamentalE}
\end{equation}
This equation is referred to as {\it the fundamental relation of equilibrium} (FRE).
Although Gibbs introduced this relation by referring to as the fundamental equation \cite{Gibbs-TD}, the name FRE is suitable by considering its meaning. By changing the variables, alternative thermodynamic potentials, namely, free energies, play an equivalent role in thermodynamics. The most familiar expression is the energy expression,
\begin{equation}
U=U(T, X_{1}, \dots, X_{M}),
\label{eq:UTeq}
\end{equation}
as presented in standard treatment \cite{Callen}.
The FRE has been derived as a logical consequence of the first and second laws (\cite{Gyftopoulos}, p.~119), which is a great advantage of GB approach. This is contrasting to Callen's approach, where the FRE is put forth as a postulation (\cite{Callen}, p.~28).

A special form of the FRE is useful for analyzing TCs.
When the Hamiltonian of a system can be diagonalized with respect to $\{ X_{j} \}$, $S$ can be factorized as
\begin{equation}
S= \sum_{j}^{M} S_{j}(X_{j}),
\label{eq:DiagonalS}
\end{equation}
where $S_{j}(X_{j})$ are the respective components. 
Although the following arguments do not require this separable form, this elegant form is convenient to see how individual TCs contribute to the total entropy. In many cases, the factorized form is expected to hold approximately at low temperatures. As the temperature decreases, the energy barriers become large compared with the thermal energy, and the TCs effectively decouple, having as independent degrees of freedom. It is well known that, at low temperatures, the harmonic approximation becomes valid. In this regime, the entropy of a crystalline solid can be written as
\begin{equation}
S = k_{\rm B} \sum_{k=1}^{M} \left[ (\bar{n}_{k}+1) \ln (\bar{n}_{k}+1) - \bar{n}_{k} \ln \bar{n}_{k} \right],
\label{eq:SofPhonons}
\end{equation}
where $M=3 N_{\rm at}-3$, and $\bar{n}_{k}$ is the Bose occupation number for the $k$-th phonon mode with the frequency $\omega_{k}$. Equation (\ref{eq:SofPhonons}) approaches zero as $T$ approaches 0.

The FRE has the following property \cite{Gibbs-TD, Callen, Gyftopoulos}.

\noindent
{\bf Corollary 5: Completeness of thermodynamic coordinates}

{\em Any thermodynamic property of a system can be obtained from the fundamental relation of equilibrium.}

\noindent
If it were not complete, the above one-to-one correspondence would not hold.

Let us investigate the significance of the present corollaries for solids.
For solids, the set $\{ \bar{\bf R}_{j} \}$ describes the equilibrium structure of a given solid. By combining Corollaries 3, 4, and 5, we reach the seemingly trivial conclusion that all thermodynamic properties of a given solid are determined by the structure. From the view of solid state physics, this conclusion is taken for granted. However, when combined with Corollary 1, this has nontrivial consequences on thermodynamics \cite{Shirai18-StateVariable}. 
Many properties of solids have history dependence: the mechanical properties of metals are affected by prior thermal history; plastic deformation is the consequence of previous mechanical loading; memory effects arise from the previously applied fields. Because such history-dependent properties are not uniquely determined solely by the current values of $T$ and $V$, they have traditionally been regarded as nonequilibrium properties. 
Now, we have seen on rigorous grounds that $\{ \bar{\bf R}_{j} \}$ are indeed the TCs. {\em Any defect state, any static state resulting from hysteresis, is an equilibrium property} as far as $\{ \bar{\bf R}_{j} \}$ do not change within the time period $\tau$ under consideration. If two glass samples with identical structures are prepared by different methods, we cannot distinguish which sample was prepared by which method. This is precisely the meaning of {\em state} in the classical thermodynamics: a state is independent of the process by which the current state was obtained. Now, we are able to answer to the fundamental question posed by Bridgman: "Which are missing variables?" \cite{Bridgman61}
The problem of history dependence is further examined in the specific heat (\cite{Shirai-SH-Liquids25}, \S 5.2) and in the thermodynamic properties of defective crystals \cite{Shirai24-hysteresis}.

It is emphasized that the present conclusion---that the equilibrium positions of atoms are TCs of solids---does not assume crystal periodicity or any regularity in atom arrangement. Therefore, this conclusion applies equally to glasses and, more generally, to all solids, including living systems. The essential point is that the equilibrium atom position of solids has the sampling invariance and have relevance with energy through the FRE. In his famous book ``What is life?", Schr\"{o}dinger described DNA as ``aperiodic crystal" followed by the explanation that ``every atom is playing there" \cite{Schrodinger44} (p.~82). The present conclusion provides the theoretical foundation for this statement. On this basis, it is desirable to revisit the definition of order. See \cite{Shirai25-OrderParams} for further details.

\subsection{Thermodynamic state space}
\label{sec:TDSS}
$M$ thermodynamic coordinates span an $M$ dimensional space. 
This space is called {\em thermodynamic state space}. (The term state space is already widely used in quantum mechanics.)

\noindent
{\bf Definition 4: Thermodynamic state space}

{\it The thermodynamic state space ${\mathscr A}$ spanned by $M$ thermodynamic coordinates, $\{ X_{j} \}$, is expressed as
\begin{equation}
{\mathscr A}=(U, X_{1}, \dots, X_{M}).
\label{eq:TDspace}
\end{equation}
}

\noindent
Here, for convenience, the internal energy $U$ is counted as the 0th coordinate and is excluded from the dimensionality of the thermodynamic state space. As noted in Introduction, the value of entropy contains an arbitrariness. We now see that this arbitrariness stems from the choice of TCs in a given problem. For example, a measurement of entropy of a Si wafer yields different values depending on whether the nuclear spin degree of freedom are included or not. Thus, the value of entropy depends on the thermodynamic state space ${\mathscr A}$ on which $S$ is expressed:
\begin{equation}
S^{\mathscr A}=S(U, X_{1}, \dots, X_{M}).
\label{eq:SinA}
\end{equation}
By expressing $S$ in the form Eq.~(\ref{eq:DiagonalS}), we see that, when entropy is evaluated in different spaces ${\mathscr A}$ and ${\mathscr B}$, one generally finds $S^{\mathscr A}(A) \neq S^{\mathscr B}(A)$, even though the two values refer to the same physical state. 

The dependence of entropy on the space on which it is expressed was also pointed out by Grandy in a different context (\cite{Grandy}, p.~150). As is often the case in information theory, however, he attributed the freedom in the entropy origin to our knowledge of TCs: if someone manipulates a new variable $X_{M+1}$ unknown to us, the value of entropy will change. 
Although the present author does not adopt the information theoretic view that entropy is not a material property, it is nevertheless a fact that its numerical value depends on the assumed space, as seen from the expressing $S^{\mathscr A}$.
The manner in which the entropy origin is fixed within this arbitrariness will be discussed in Sec.~\ref{sec:Frozen-coordinates}. Before proceeding there, further clarification of the properties of thermodynamic state space is required.

Suppose that we have two entropy values of $S^{\mathscr A}(A)$ of state $A$ evaluated in a thermodynamic state space ${\mathscr A}$ and $S^{\mathscr B}(B)$ of state $B$ in another space ${\mathscr B}$.

\noindent
{\bf Corollary 6: Comparability condition}

{\em In order to make comparison of entropies of $S^{\mathscr A}(A)$ and $S^{\mathscr B}(B)$ possible, the two thermodynamic state spaces ${\mathscr A}$ and ${\mathscr B}$ must have the same dimensions.}

\noindent
When the two states, $A$ and $B$, are evaluated on the same space, it is always possible to find a reversible path in Eq.~(\ref{eq:defS}), because the state can be  continuously changed from $A$ to $B$ {\em without breaking constraints}. Thus, entropy difference $\Delta S_{AB}$ is well measured and discussed. Otherwise, entropy increase/decrease should not be discussed by just looking at the difference $\Delta S_{AB}$.

Let us consider the thermodynamic state space of solids more specifically. For solids, the thermodynamic state space is expressed as
\begin{equation}
{\mathscr A}=(U, \{ \bar{\bf R}_{j} \} ),
\label{eq:TDspace1}
\end{equation}
which has $3 N_{\rm at}$ dimensions. 
Although an atom can change its equilibrium position $\bar{\bf R}_{j}$ from a regular position $\bar{\bf R}_{j}^{1}$ by forming defect states, the equilibrium position $\bar{\bf R}_{j}$ cannot be a continuous variable. As shown in Fig.~\ref{fig:meta-frozen}, atoms in a solid are so strongly localized that local minima in the potential energy are well separated by energy barriers. There exists a large number of distinct atom configurations, which are labelled by an index $K$. For solids, it is more convenient to partition the full thermodynamic space $\{ \bar{\bf R}_{j} \}$ into discrete sets $\{ \bar{\bf R}_{j}^{K} \}$, each corresponding to a particular equilibrium configuration $K$. 
The state of the perfect crystal is represented by a single point $\{ \bar{\bf R}_{j}^{1} \}$ in the thermodynamic state space $\{ \bar{\bf R}_{j} \}$. Any defect state of the crystal is represented by another point $\{ \bar{\bf R}_{j}^{K} \}$. Until this point, the term {\it configuration} was used in a empirical manner, but from now on it is used in a rigorous manner to refer to an equilibrium configuration. 

\noindent
{\bf Definition 5: Configuration}

{\em A configuration is specified by a set of equilibrium positions of atoms $\{ \bar{\bf R}_{j} \}$. } 

\noindent
When there are $N_{\rm c}$ different configurations, the thermodynamic state space is expressed as:
\begin{equation}
{\mathscr A}=(U, \{ \bar{\bf R}_{j}^{(K)} \}  ),
\label{eq:TDSDiscrete}
\end{equation}
where $(K)$ enumerates all configurations $K =1, \dots, N_{\rm c}$. A set $\{ \bar{\bf R}_{j}^{K} \}$ may be regarded as a coordinate transformation $i \leftrightarrow K$, similar to the normal modes transformation.
Since a configuration is specified by equilibrium atom positions $\{ \bar{\bf R}_{j}^{K} \}$, the term configuration is synonym with structure. At fixed $T$, different states of a solid therefore correspond to different structures. Again notice that any regularity or symmetry in atomic arrangement has no importance for the present term of structure. The essential feature of both structure and configuration is the sampling invariance at equilibrium \cite{Shirai25-OrderParams}. Hence, we are able to speak of the structure for glasses and DNA on the same basis for crystals.

Since a configuration is specified by TCs $\{ \bar{\bf R}_{j}^{K} \}$, it has a well-defined meaning only for solids. However, it is sometimes convenient to extend the concept of configuration to liquids, as discussed in Sec.~\ref{sec:glass-transition}. For a liquid, one may consider an instantaneous configuration defined by $K(t) = \{ {\bf R}_{j}(t) \}$. When describing the transition from solid to liquid, we can employ $K(t_{m})$ at the time $t_{m}$ of melting as the configuration of the liquid. In this manner, an analytic continuation from the solid phase to the liquid can be achieved. However, this extension is used only for such a limited purpose.

\subsection{Frozen and active coordinates}
\label{sec:Frozen-coordinates}
The most important quantity in thermodynamics is energy. The concept of {\it thermodynamic relevance}, as introduced in Sec.~\ref{sec:TCgeneral}, plays a role through its relation with energy. Even a seemingly trivial coordinate, such as the center of mass in a gas, can become a genuine TC, when it is energetically relevant.
When Kline and Koenig studied the definition of a state variable, they considered the following example \cite{Kline57,Hatsopoulos}. The properties of water in a container do not change when measured at different elevations $H$ at which the container is held; they depend only on $T$ and $V$. The FRE is expressed by these two coordinates, $(T, V) \equiv \{ X_{j}^{0}\}$. 
However, there are cases in which $H$ does affect the properties of water. In a hydraulic power plant, the elevation $H$ is the most important coordinate, and the FRE depends on $H$.
If the gate of the upper reservoir is closed so that thermal communication between the upper and lower water is prevented, $H$ no longer enters the FRE. In this case, $H$ is called a frozen coordinate.

\noindent
{\bf Definition 6: Frozen coordinates}

{\em Frozen coordinates are the thermodynamic coordinates that are irrelevant to the present scope of the problem}. 

\noindent
Let us denote frozen coordinate $X$ by $\hat{X}$. Then, the FRE is expressed as $S(\{ X_{j}^{0}\}; \hat{H}) = S_{X}(\{ X_{j}^{0} \}) + S_{H}(\hat{H})$. The part $S_{H}(\hat{H})$ is constant. The role of the constraint in this case is to completely inhibit $\hat{H}$ from varying. When the gate is opened, the upper water falls into the lower reservoir, and $H$ begins to vary and becomes a real variable. Now, $H$ must be taken into account in $S$, as $S=S(\{ X_{j}^{0}\}, H)$. The gravitational potential of the upper water is converted into the thermal energy of the lower water. This is the process where the origin of entropy is reconstructed from one to another space. Because the constraint is broken, this process is an irreversible process. 

The entropy of a perfect crystal is zero at $T=0$. Nevertheless, a certain disorder---for example, the random distribution of isotopes---may still exist within this apparent perfection. This isotopic randomness does not affect the thermodynamic properties of this crystal under ordinary conditions; therefore, the isotopes distribution constitutes a frozen coordinate. However, when an isotope enrichment process is investigated, the distribution of isotopes becomes a genuine variable, and therefore must be included in the state space. 
There are infinitely many possible frozen coordinates \cite{Grad61,Jaynes65}. The anthropomorphic feature of entropy originates from the presence of such frozen coordinates. A frozen coordinate has no relevance to the current thermodynamic properties within the present scope of interest. Therefore, a frozen coordinate has no $T$ dependence in thermodynamic properties and is thus thermodynamically inactive. 

\noindent
{\bf Definition 7: Active coordinates}

{\em Non-frozen coordinates are active coordinates.}

\noindent
Any TC appearing in the FRE is an active coordinate, since it influences other variables through the FRE. An active coordinate introduces temperature dependence into its thermodynamic properties via Eq.~(\ref{eq:UTeq}). For solids, the active coordinate $\bar{\bf R}_{j}$ is associated with the phonon component ${\bf u}_{j}(t)$, through ${\bf R}_{j}(t) =\bar{\bf R}_{j} + {\bf u}_{j}(t)$. This component introduces the $T$ dependence of $S(\bar{\bf R}_{j})$, as expressed in Eq.~(\ref{eq:SofPhonons}). In contrast, frozen coordinates exhibit no temperature dependence and remain inactive in thermodynamic responses.

When a thermodynamic state space ${\mathscr A}=(U, \{ X_{j}\} )$ of a material is a subspace of ${\mathscr B}=(U, \{ X_{j}\}, \{ Y_{j}\} )$ of the same material, we often need to compare the entropy $S^{\mathscr A}(A)$ of state $A$ in the space ${\mathscr A}$ to the entropy $S^{\mathscr B}(B)$ of state $B$ in the space ${\mathscr B}$.
From Corollary 6, a comparison of entropy is justified only when the two spaces have the same dimensionality. This requirement can be satisfied by patching the space ${\mathscr A}$ with the frozen coordinates $\{ \hat{Y}_{j} \}$, as $(U, \{ X_{j}\}; \{ \hat{Y}_{j}\} )$. In physical problems, the values $\{ \hat{Y}_{j} \}$ must always exist.
This is because the subspace $( \{ Y_{j}\} )$ corresponds to physically meaningful degrees of freedom. Therefore, it must be possible to identify a state $A$ within the full space that includes $( \{ Y_{j}\} )$: the previous value $S^{\mathscr A}(A)$ is merely the result of having ignored the existence of $( \{ Y_{j}\} )$.

\noindent
{\bf Corollary 7: Extending thermodynamic state space}

{\em It is possible to extend the thermodynamic state space ${\mathscr A}=(U, \{ X_{j}\} )$ by adding frozen coordinates $\{ \hat{Y_{j}}\}$ to form ${\mathscr B}=(U, \{ X_{j}\}; \{ Y_{j}\} )$.}

\noindent
Under this extension, the entropy of $A$ transforms as
\begin{equation}
S^{\mathscr B}(A) = S^{\mathscr A}(A)+S_{0},
\label{eq:SBeqSAplus0}
\end{equation}
where $S_{0} = S(\{ \hat{Y}_{j} \} )$ is a constant. The comparison between $S^{\mathscr B}(A)$ and $S^{\mathscr B}(B)$ is now well defined, since both entropies are expressed in the same state space. The additive feature in Eq.~(\ref{eq:SBeqSAplus0}) follows from the definition of frozen coordinates: the constraint on $Y_{j}$ is sufficiently strong that the change in $Y_{j}$ is completely decoupled to other TCs $\{ X_{j} \}$. 
If the constraint fixing the value $\hat{Y}_{j}$ of $j$th component is adiabatically removed, a spontaneous change is induced unless the initial state happens to be in equilibrium. The system then relaxes to a new equilibrium state $A'$, characterized by a different value $Y_{j}'$. Because thes adiabatic removal of a constraint is an irreversible process, entropy is produced: $S_{\rm prod}=S(Y_{j}')-S(\hat{Y}_{j})$. 

Let us consider the special role of frozen coordinates in solids. In this context, it is essential to recognize that the TCs are not continuous variables. The equilibrium positions $\bar{\bf R}_{j}$ form a discrete set. The equilibrium states of a solid are represented by discrete points in the $3 N_{\rm at}$ dimensional space ${\mathscr A}=(U, \{ \bar{\bf R}_{j}^{K} \} )$. Any two configurations are separated by the energy barrier. As $T$ decreases, transitions between them become increasingly inhibited. At sufficiently low $T$, the solid becomes trapped in a certain configuration, say, $K=1$. Only $\{ \bar{\bf R}_{j}^{1} \}$ remain active coordinates: they have phonon contributions. 
Let us denote the thermodynamic state space spanned by the active coordinates as ${\mathscr A} = (U, \{ \bar{\bf R}_{j}^{1} \})$.

\noindent
{\bf Definition 8: Active configuration}

{\em The active configuration consists solely of the active coordinates $\{ \bar{\bf R}_{j}^{1} \}$. }

\noindent
The remaining configurations are frozen configurations. The full thermodynamic state space is written as
\begin{equation}
{\mathscr B}=(U, \{ \bar{\bf R}_{j}^{1} \}; \{ \hat{\bf R}_{j}^{(K')} \} ),
\label{eq:TDS-FC1}
\end{equation}
where $\{ \hat{\bf R}_{j}^{(K')} \} = \{ \hat{\bf R}_{j}^{K} \}_{K = 2, \dots, N_{\rm c} }$.
We can compare the entropy of state $A$ in ${\mathscr A} = (U, \{ \bar{\bf R}_{j}^{1} \}) $ to the entropy of $B$ in ${\mathscr B}$ by patching frozen coordinates $\{ \hat{\bf R}_{j}^{(K')} \}$.
When the constraints that fix $\{ \hat{\bf R}_{j}^{(K')} \}$ are adiabatically removed, a spontaneous change is induced. Since the frozen coordinates have no phonon components, their contribution to entropy is purely configurational entropy, $S_{\rm conf}=k_{\rm B} \ln N_{\rm c}$, and independent of independent. Here, for simplicity, we assumed that all the configurations are energetically degenerate. Upon removal of the constraints, the system can access the previously frozen configurations. Because this process is irreversible, entropy production occurs, and its magnitude is $S_{\rm prod}=S_{\rm conf}$. 

It should be emphasized that the distinction between active and frozen is not at our disposal but must be determined experimentally. Frozen coordinates do not enter in the relation between $U$ and the active coordinates. Consequently, thermodynamic measurements---such as specific heat---do not detect the frozen coordinates.
Different samples of silicon possess different defect structures. By Definition 8, the active configuration of a given sample is precisely its present defect structure. For that sample, the structures of all other samples are frozen configurations. Indeed, the measured specific heat of the given sample reflects the phonon spectrum of the current defect structure and is unaffected by alternative defect structures. Similarly, for a given sample of glass, the current structure of the sample constitutes the active configuration, while all other possible configurations are frozen configurations.

\subsection{Meaning of ensemble average}
\label{sec:ensemble}
Someone says that ambiguity of entropy origin can be removed when statistical mechanics is used to calculate entropy. This is not true. Ambiguity of entropy cannot be eliminated even in statistical mechanics without clarifying the meaning equilibrium.
In statistical mechanics, there are two ways of defining entropy, which have given rise to a long-standing debate \cite{Jaynes65,Penrose79, Lebowitz93,Goldstein04,Uffink01,Callender04}. 
Boltzmann entropy is defined as
\begin{equation}
S_{\rm B}(A) = k_{\rm B} \ln \Gamma_{A}, 
\label{eq:Boltzmann-S}
\end{equation}
for a macroscopic state $A$, where $\Gamma_{A}$ is the number of microscopic states corresponding to $A$.
Gibbs entropy is given by
\begin{equation}
S_{\rm G} = -k_{\rm B} \sum_{i} p_{i} \ln p_{i}, 
\label{eq:Gibbs-S}
\end{equation}
where $p_{i}$ is the probability that the microscopic state $i$ occurs. Here, the summation is taken over all states in the ensemble space ${\cal M}$. Both definitions yield the same value when there is only a single macroscopic state $A$; in this case, $p_{i} = 1/\Gamma_{A}$, and hence $S_{\rm G} = S_{\rm B}$. This situation applies to ideal gases and perfect crystals. Generally, however, they give different results. The crucial issue, therefore, is the extent to which sampling is taken into account for the ensemble space ${\cal M}$.

This contrasting character becomes clearer when the entropy of a solid is studied. 
The ensemble space ${\cal M}$ consists of all accessible configurations, $\{ K \}$. However, whether a given state is regarded as accessible depends on the scope of problem under consideration.
For example, when the specific heat of diamond is studied, only the perfect structure $K=1$ is sufficient in calculating Eq.~(\ref{eq:Gibbs-S}), even though there is no perfect crystal. By contrast, if one is interested in calculating the vacancy concentration, configurations containing vacancies, $K \neq 1$, must be taken into account. If one investigates the phase transition between diamond and graphite at high pressures and high temperatures, the graphite structure must also be included among the members of ${\cal M}$. In this way, physicists choose an appropriate ensemble space according to the physical intuitions. 

Despite this, entropy is a state function in a similar way as energy is (see just after Corollary 4). According to the rigorous definition of equilibrium adopted in the present study, only a single configuration $K=1$ qualifies as the true equilibrium state; hence, the ensemble average should not be performed. Configurations other than $K=1$ are frozen coordinates. For example, in a crystal containing vacancies, each sample possesses only one particular vacancy configuration, and hence the configuration entropy $S_{c}$ vanishes. The question then arises: why do we ordinarily use $S_{c} = k_{\rm B} \ln W_{c}$ for this case? This expression becomes necessary when one investigates the equilibrium concentration of vacancies, $n_{d}$. In that context, it is tacitly assumed that the host crystal ($X$) and vacancy ($V$) are in chemical equilibrium expressed by
\begin{equation}
X \leftrightarrow V + X_{\rm atom}, 
\label{eq:form-V}
\end{equation}
where $X_{\rm atom}$ denotes an isolated atomic state of $X$. A chemical equilibrium is established at the condition that the forward and backward reactions occur equally.
Such equilibrium is realized near the melting temperature. Under these conditions, all vacancy configurations are accessed within the experimental timescale and therefore must be sampled. A particular defect configuration cannot by itself constitute an equilibrium state. As discussed in Sec.~\ref{sec:constraint}, the timescale required to reach equilibrium plays a critical role in determining whether a state can be regarded as equilibrium. 

A problem of measuring entropy in the above case is irreversibility. Since mixing is usually an irreversible process, Eq.~(\ref{eq:defS}) cannot be directly applied to entropy measurement. Therefore, in experiment, the mixing process must be replaced by a reversible one, for example by semipermeable membranes \cite{Cengel} or by electrochemical method \cite{Eastman33}. In theoretical side, It is quite difficult to treat irreversible processes by ensemble average.

\section{The refinement of the third law expression}
\label{sec:third-law}
By establishing rigorous definitions of equilibrium and TC, we are not in position to construct a rigorous statement of the third law. All existing structures of solids, including glasses, are equilibrium states within their respective relaxation times (Corollary 1). This statement applied in particular at $T=0$ (Corollary 2). These states are specified by TCs $\{ \bar{\bf R}_{j} \}$ (Corollary 3). Moreover, a full set of $\{ \bar{\bf R}_{j} \}$ uniquely determines the equilibrium state of a solid (Corollaries 4 and 5). 
A distinctive feature of solids is that the equilibrium atomic positions are not continuous variables. Hence, use of configuration index $K$ is more convenient for representing TCs of a solid. A given sample of a solid has its own structure by definition. This structure constitutes the active configuration of that sample (Definition 8). 
Let us denote this active configuration of a given sample by $K=1$. The corresponding thermodynamic state space is ${\mathscr A}=(U, \{ \bar{\bf R}_{j}^{1} \})$. The full thermodynamic state space is expressed as $(U, \{ \bar{\bf R}_{j}^{1} \}; \{ \hat{\bf R}_{j}^{(K')} \} )$. At sufficiently low temperatures, the configuration $K=1$ is only the equilibrium configuration. 

\noindent
{\bf Postulation 2: The third law of thermodynamics}

{\em Entropy $S^{\mathscr A}$ of any configuration of a solid vanishes as $T$ approaches zero, provided that the entropy is evaluated in the thermodynamic state space ${\mathscr A}$ that has only one active configuration, ${\mathscr A}=(U, \{ \bar{\bf R}_{j}^{1} \})$.} 

\noindent
Because entropy is a state function (Corollary 4), only this way of assigning the entropy value at $T=0$ is permissible.
In this manner, the third law attains the status of a universal law without exception. The essential problem with the previous theories is that they did not recognize that a solid possesses many equilibrium states, each of which has its own value of entropy. 
In the new theory, it is a meaningless question to ask what entropy of the ``generic" silicon is, because different wafers of silicon contain different defects and therefore have different entropies. An ensemble average over different samples does not make sense, as discussed in Sec.~\ref{sec:ensemble}. In the same vein, any sample of silica glass is in equilibrium, and the third law states that all {\rm samples} of silica glass have zero entropy at $T=0$.

When frozen coordinates are taken as TCs, the thermodynamic state space is extended to ${\mathscr B}=(U, \{ \bar{\bf R}_{j}^{1} \}; \{ \hat{\bf R}_{j}^{(K')} \} )$, and the entropy $S^{\mathscr B}$ takes a finite value even at $T=0$.
This nonzero value is the residual entropy and is given by
\begin{equation}
S_{\rm res} = S^{\mathscr B}(\{ \hat{\bf R}_{j}^{(K')} \} ).
\label{eq:ThirdLaw3}
\end{equation}
All previous calculations showing residual entropies are based on the configurational contribution of entropy without making distinction between active and frozen coordinates.
A question remains as to why the residual entropy is observed in experiments. As explained in the Kivelson and Reiss debate in Introduction, it is certain that the residual entropy observed in experiments is a real quantity, in the sense that $S_{\rm res}$ is detected within experimental accuracy. The mechanism of detecting $S_{\rm res}$ depends on the specific measurement method---whether calorimetric or electrochemical---and on the types of disorder involved, such as random alloys or glasses. The author does not attempt to explain all such cases here. Instead, it suffices to illustrate the basic mechanism by considering the glass transition measured by calorimetry as an example. For further details of the Kivelson and Reiss debate, see Sec.~\ref{sec:glass-transition}.

In the specific heat measurement, the residual entropy $S_{\rm res}$ of glass is obtained by cooling the system from the liquid phase. In the liquid state, all configurations are observed during the measurement; thus, the entropy of the liquid state $A$ at a high temperature $T_{A}$ is evaluated in the full space ${\mathscr B}$, resulting in $S^{\mathscr B}(A)$. Since the specific heat measurement is assumed to carry out along a reversible path in Eq.~(\ref{eq:defS}), the thermodynamic state space does not change from ${\mathscr B}$ throughout the measurement. Consequently, the entropy of the glass state $B$ at a low temperature $T_{B}$, as determined by specific heat measurements, is evaluated in the same space ${\mathscr B}$ and is given by
\begin{equation}
S^{\mathscr B}(B) = \int_{T_{A}}^{T_{B}} \frac{C_{p}(T)}{T} d T + S^{\mathscr B}(A),
\label{eq:CintegT}
\end{equation}
where $C_{p}$ is the isobaric specific heat of the substance. The residual entropy is contained in the integration constant $S^{\mathscr B}(A)$. The entropy of liquid $S^{\mathscr B}(A)$ was beforehand obtained by another specific heat measurement performed during a heating process from the crystalline state. The residual-entropy contribution included in $S^{\mathscr B}(A)$ is not cancelled by the first term on the right-hand side of Eq.~(\ref{eq:CintegT}). As a result, $S_{\rm res}$ remains finite even at $T=0$. This is the origin of the experimentally observed residual entropy. We should not interpret the finite value $S^{\mathscr B}(B) \neq 0$ as implying $S^{\mathscr A}(B) \neq 0$. For ordinary crystals, the entropy value of a crystal obtained by specific heat measurement is evaluated only in the active space, namely, $S^{\mathscr A}(B)$. This implies that the contraction of the thermodynamic state space from the full space ${\mathscr B}$ to the subspace ${\mathscr A}$ must occur at the transition. This contraction takes place reversibly and is accompanied by an energy exchange in the form of latent heat.

The following remarks are worth mentioning in order to emphasize the significance of this new expression of the third law.
\begin{enumerate}
\item{The new expression is deduced from the investigation on equilibrium, which shows that the current states of all existing materials are equilibrium states, at $T=0$.}
This conclusion frees us from the previously unsupported assumption that disordered materials will eventually transform into ordered state if given sufficiently long waiting time. At $T=0$, the current states of glasses, random alloys, and defect-containing solids are already equilibrium states at present. The new expression eliminates speculative expectations about hypothetical future events.

\item{No need arises to determine which state is the lowest-energy state or to distinguish metastable states from the stable one.} A conventional explanation for the residual entropy of disordered materials assumes that the lowest-energy state is an ordered state and, accordingly, that the disordered states are not equilibrium states. This view leads to the conclusion that the thermodynamics laws cannot be applied to nonequilibrium states. 
A logical difficulty in this conventional explanation is that the validity of the third law then depends entirely on the hypothesis that the order state is the lowest-energy state. In that case, this hypothesis would have to be elevated to the status of a thermodynamic law, and the present third law would be reduced to a derived consequence.
Recent experimental results disprove this hypothesis. For boron crystal, it has been demonstrated that the energy of the disordered structures is lower than that of the ordered structure \cite{Ogitsu09}. Similar findings have been obtained for several boron compounds \cite{Xie15, Rasim18}.

\item{The no-exception features described in (1) and (2) establish the third law as a truly a universal law of physics. Accordingly, the expression is fully compatible with the universal validity of the unattainability principle of the absolute zero temperature. The logical equivalence of these two statements has been demonstrated in another paper \cite{Shirai18-Unattain}.
}
\end{enumerate}

\section{Arguments on the glass transition}
\label{sec:glass-transition}
Residual entropy essentially arises from the irreversibility when a constraint is removed. Irreversibility in solids are highly complicated and constitutes one of the most difficult problems in thermodynamics. In experiment, any real process involves irreversibilities more or less, which obscure our understanding of the microscopic origin of residual entropy. A notable example is the debate initiated by Kivelson and others \cite{Kivelson99,Speedy99,Moller06,Mauro07,Gupta07,Goldstein08,Reiss09,Gupta09,Aji10}. It is therefore worth to analyze this debate here.
In Sec.~\ref{sec:third-law}, we have seen that the entropy of a glass vanishes at $T=0$. This conclusion agrees with the conclusion of Kivelson and Reiss claiming zero entropy \cite{Kivelson99}. On the other hand, the present theory also permits the reality of residual entropy. How the two seemingly contradictory views are reconciled must be explained. 
The real processes of the glass transition are irreversible due to atom relaxation, hysteresis, and other dissipative effects, which have led to considerable confusion in entropy evaluation \cite{Langer88}. Thermodynamic analysis of hysteresis itself requires a comprehensive treatment \cite{Shirai24-hysteresis}. Although hysteresis influences the evaluation of residual entropy, the residual entropy remains even if the hysteresis is eliminated \cite{Moller06}. Thus, hysteresis is not essential to the origin of residual entropy \cite{Bestul65,Goldstein08}, and hence, in the following, we ignore any effect caused by the hysteresis in order to concentrate on the principal mechanism responsible for residual entropy. Any frictional forces that generate additional irreversibility are likewise neglected. 

Let us consider a specific heat measurement of a glass during a heating process ($a$) from $T=0$ to $T_{m}+$, namely, slightly above the melting temperature $T_{m}$. Here, specific heat is always inferred as isobaric specific heat $C_{p}$; therefore, the subscript $p$ will be omitted in the following. During process ($a$), the glass substance undergoes the sequence of transformations: (glass) $\rightarrow$ (supercooled liquid) $\rightarrow$ (normal liquid).
Upon heating, the specific heat of the glass, $C_{\rm gl}(T)$, increases. At the glass-transition temperature $T_{g}$, the glass transforms into the supercooled liquid, accompanied by a jump $\Delta C_{\rm gl}$ in the specific heat \cite{Shirai22-SH,Shirai22-Silica}. The glass transition, however, has a finite width $\Delta T_{g}=T_{g2}-T_{g1}$: this interval is called the transition region \cite{Moynihan74}. The finite width $\Delta T_{g}$ spreads the latent heat over this temperature interval, appearing as the apparent heat contribution in the specific heat. The specific heat in this region is denoted by $C_{\rm tr}$. Upon further heating, the supercooled liquid transforms into the normal liquid at $T_{m}$. The specific heat of the supercooled liquid, $C_{\rm sl}$, is generally larger than that of crystal, $C_{\rm cr}$: their difference is known as the excess specific heat, $C_{\rm ex}=C_{\rm sl}-C_{\rm cr}$. The specific heat of the normal-state liquid is denoted as $C_{\rm li}$. The same subscripts are also used to denote the corresponding entropies $S$.

Along heat path ($a$), the change in entropy from $T=0$ to $T_{m}+$ is obtained by integrating $C$ as
\begin{equation}
S_{\rm li}(T_{m}+) - S_{\rm gl}(0) =  \int_{0}^{T_{g1}} \frac{C_{\rm gl}(T) }{T} dT 
+ \int_{T_{g1}}^{T_{g2}} \frac{C_{\rm tr}(T) }{T} dT
+\int_{T_{g2}}^{T_{m}+} \frac{C_{\rm sl}(T) }{T} dT,
\label{eq:DeltaS-a}
\end{equation}
where the difference between $C_{\rm sl}$ and $C_{\rm li}$ above $T_{m}$ is neglected. Previous experiments have deduce a nonzero value of $S_{\rm gl}(0)$ ($> 0$) from Eq.~(\ref{eq:DeltaS-a}). In contrast, Kivelson and Reiss claimed that $S_{\rm gl}(0) = 0$ \cite{Kivelson99}. They are therefore requested to explain what is incorrect in Eq.~(\ref{eq:DeltaS-a}).
Since calorimetry is a well-established method for solids and liquids, respectively, the first and the last terms in the right-hand side of Eq.~(\ref{eq:DeltaS-a}) are reliable. Also the value $S_{\rm gl}(T_{m}+)$ must be reliable.
The only conceivable source of error leading to the observed residual entropy would therefore lie in the second term \cite{Kivelson99}. Kivelson and Reiss attributed this putative error to the irreversibility of the glass transition process. However, as already studied by others \cite{Bestul65,Goldstein08,Aji10}, the effect of irreversibility during the glass transition can be suppressed at minimum by careful measurement. Even though the irreversibility is removed at a level to give no significant contribution to the second integration term, the residual entropy remains. 
\begin{figure}[ht!]
\centering
\includegraphics[width=140mm, bb=0 0 768 450]{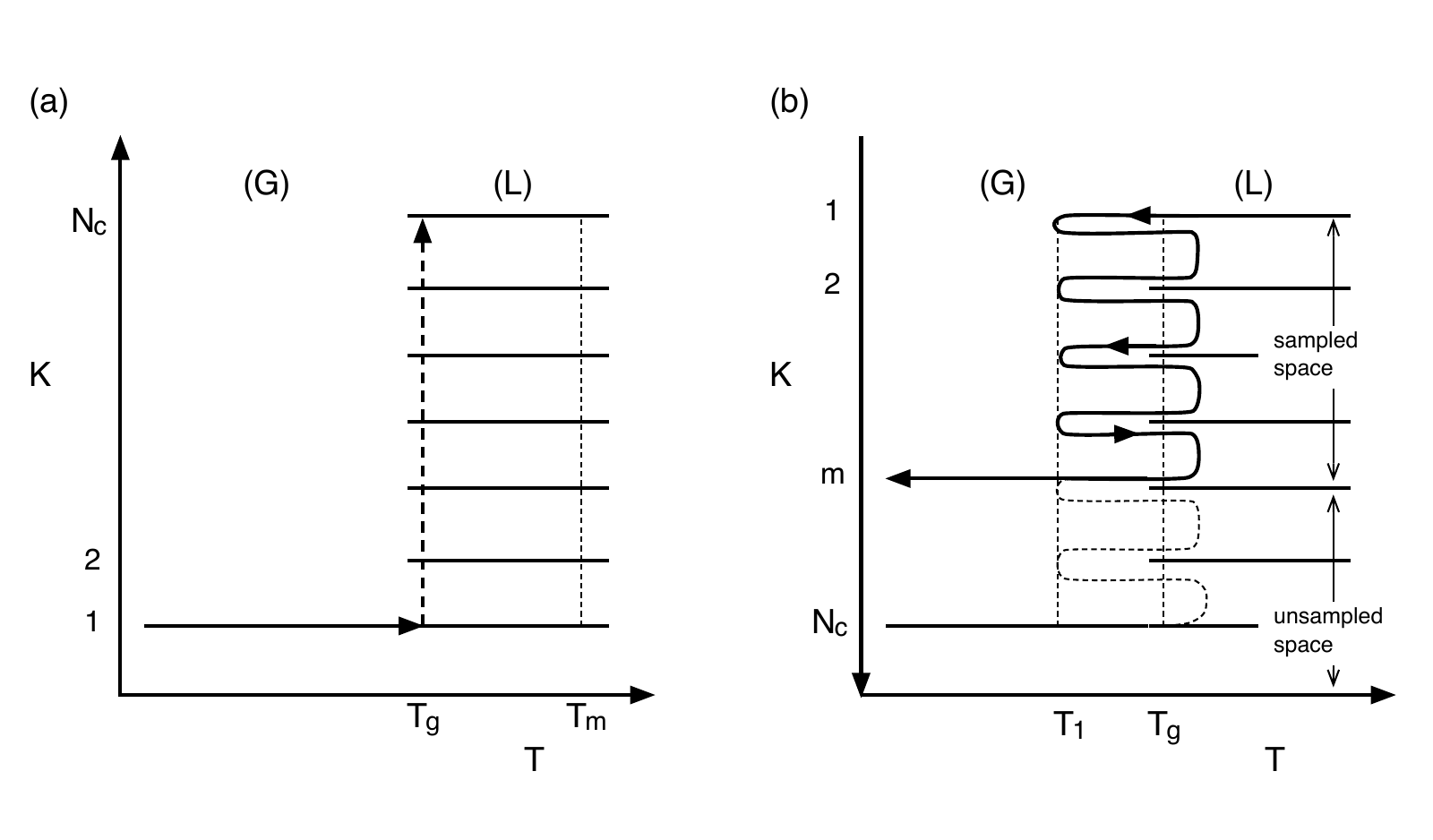}
\caption{
Glass transition represented in the configuration space $K$ versus $T$: (a) heating process starting from the glass state, for which $S_{\rm gl}^{\mathscr A}(0)=0$, (b) cooling process starting from the liquid state at a reference temperature $T_{m}+$, at which the entropy is $S_{\rm li}^{\mathscr B}(T_{m}+)$. The cooling process consists of many cooling and heating steps between $T_{g}+$ and $T_{1}$. The thick dashed line with an arrow indicates an irreversible process, and the thick solid lines with arrows indicate reversible processes.}
\label{fig:AtoB} 
\end{figure}

This seemingly contradicting issue can be resolved by making the distinction between active and frozen coordinates, as described in Sec.~\ref{sec:third-law}.
The structure of the thermodynamic state space relevant to the glass transition is schematically illustrated in Fig.~\ref{fig:AtoB} (a). In this representation, the thermodynamic state space of liquid is partitioned by discrete configurations $K$. However, readers should be noted that this was done only for practical convenience, as explained at the end of Sec.~\ref{sec:TDSS}. Furthermore, the transition width $\Delta T_{g}$ is neglected in this scheme for simplicity.

Heating process ($a$) begins from the glass state at $T=0$. The present theory dictates that $S_{\rm gl}(0)=0$. The corresponding thermodynamic state space consists only of the active configuration $K=1$, that is, ${\mathscr A}=(U, \{ \bar{\bf R}_{j}^{1} \})$. The measurement proceeds along the line $K=1$ until reaching at $T_{g}$. Above $T_{g}$, whether in the supercooled or in the normal liquids, the thermodynamic state space must be the full space, ${\mathscr B}=(U, \{ \bar{\bf R}_{j}^{(K)}\} )$, consisting of all configurations, since all atoms in the liquid are free to explore the entire configuration space. Consequently, the thermodynamic state space must change from ${\mathscr A}$ to ${\mathscr B}$ at $T_{g}$, 
\begin{equation}
(U, \{ \bar{\bf R}_{j}^{1} \}; \{ \hat{\bf R}_{j}^{(K')} \} ) \rightarrow (U, \{ \bar{\bf R}_{j}^{(K)}\} ).
\label{eq:trasAtoB}
\end{equation}
This indicates that the previously frozen coordinates $\{ \hat{\bf R}_{j}^{(K')} \}$ becomes fully activated on melting: the corresponding constraints $\xi(\{ \hat{\bf R}_{j}^{(K')} \})$ have been removed. Removing the constraints is an irreversible process, and accordingly, an entropy production $S_{\rm prod}$ occurs without heat exchange: that is, an uncompensated entropy increase takes place. This situation is analogous to the entropy increase during the free expansion of gas: a gas initially confined to one side of a container that is separated by an internal wall expands by quickly removing the wall. In a similar sense, the glass transition viewed from the low temperature side can be looked upon as a free expansion of the configuration space accessible to the atoms. The resulting uncompensated entropy increase corresponds to the configuration entropy $S_{\rm conf}$, which turns to the residual entropy $S_{\rm res}$ given by Eq.~(\ref{eq:ThirdLaw3}).
Therefore, the entropy increase by $S_{\rm res}$ is captured by the specific heat measurement.
This argument does not imply that the specific heat measurement in the transition region is incorrect. The physics does not change. The problem lies in our interpretation: we failed to recognize the change in the entropy origin associated with the change in the thermodynamic state space from ${\mathscr A}$ to ${\mathscr B}$. The apparent discrepancy arises from this unrecognized change in the underlying state space.

When the glass transition is observed during cooling, a different type of difficulty arises.
Consider a cooling process ($b$) from liquid state: (normal liquid) $\rightarrow$ (supercooled liquid) $\rightarrow$ (glass). The entropy of the glass at $T=0$ is obtained by
\begin{equation}
S_{\rm gl}(0) =  \int_{T_{m}+}^{0} \frac{C(T) }{T} dT 
  +S_{\rm li}(T_{m}+),
\label{eq:DeltaS-b}
\end{equation}
where distinct regions of integrations are taken together to a single region from $T_{m}+$ to $0$. In this expression, the initial entropy $S_{\rm li}(T_{m}+)$ is evaluated in the thermodynamic state space ${\mathscr B}=(U, \{ \bar{\bf R}_{j}^{(K)}\} )$. Hence, $S_{\rm li}^{\mathscr B}(T_{m}+)$ contains the component of configurations, $S_{\rm conf}$. 
The integration in Eq.~(\ref{eq:DeltaS-b}) assumes that this cooling process is reversible. When no heat exchange except cooling or no work exchange occurs, this reversibility requires that the same thermodynamic state space ${\mathscr B}$ must retain throughout the entire cooling path ($b$). Therefore, the value $S_{\rm gl}(0)$ obtained by Eq.~(\ref{eq:DeltaS-b}) is the value evaluated in the space ${\mathscr B}$, that is, $S_{\rm gl}(0) = S_{\rm gl}^{\mathscr B}(0)$.
Thus, we see that the residual entropy is given by the imbalance between the integration part and the integration constant $S_{\rm li}^{\mathscr B}(T_{m}+)$ in the right-hand side of Eq.~(\ref{eq:DeltaS-b}).
Once it is recognized that the experimentally observed quantity $S_{\rm gl}^{\mathscr B}(0)$ is a different quantity from the quantity $S_{\rm gl}^{\mathscr A}(0)$ that third law refers to, the apparent contraction is resolved.

However, a conceptual difficulty still remains. Even though the nonzero value $S_{\rm gl}^{\mathscr B}(0)$ is observed experimentally, the measured sample somehow selected a particular configuration, say $K=1$, only. The nonzero $S_{\rm gl}^{\mathscr B}(0)$ indicates that the entropy of this sample contains information of other configurations. How can this sample ``know" about configurations of other samples? This issue of causality was the original question raised by Kivelson and Reiss \cite{Kivelson99}. 

Consider interrupting the cooling process at a temperature $T_{1}$ slightly below $T_{g}$. Even though $T_{1}$ is very close to $T_{g}$, this glass sample at this stage has already selected and fixed a particular configuration, denoted by $K_{1}$. According to the present theory, all other configurations $K' \ (\neq K_{1})$ become irrelevant to the present state, and hence the entropy must decrease by the amount $S_{\rm conf}$. This decrease must be compensated by an increase in $S$ somewhere; otherwise, the second law would be violated \cite{comment5}. For glasses, no latent heat is generated, and thus the second law seems to break.
Mauro and Gupta explained this fundamental difficulty as an intervention of external constraint: the constraint in this case is, in their interpretation, the finite time of observation by human \cite{Mauro07,Gupta07}. This view seeks the resolution outside thermodynamics, making it difficult to regard a desirable explanation.
There exists another interpretation. When Eastman and Milner first demonstrated the residual entropy of a random alloy, they explained that, although the configuration of a given glass sample is fixed, we do not know which one is realized among many configurations \cite{Eastman33}. This explanation is partly correct, but it carriers a subjective tone. Indeed, it naturally provokes a further question (see, for example, \cite{Sethna}, p.~86): if we were to determine the exact structure of this sample, would $S_{\rm res}$ vanish? Whether entropy depends on our knowledge is the central contention in the long-standing debate between information theory and thermodynamics \cite{Leff-Rex2,Jaynes79,Denbigh81}. The present theory adopt the traditional standpoint of thermodynamics that entropy is a property of material: at least, the entropy in thermodynamics must be so.

The key to the resolution again lies in a proper understanding of equilibrium. 
When entropy change is measured by calorimetry, the path must be reversible, as required by Eq.~(\ref{eq:defS}). A reversible process requires quasi-static evolution, although it is not sufficient \cite{Zemansky}. A quasi-static process consists of a succession of equilibrium states, each differing from the previous one by an infinitesimal change. 
Let us apply this principle to the present cooling problem. Figure \ref{fig:AtoB} (b) schematically illustrates how a reversible cooling process can be constructed for this problem. Recall that, in a chemical equilibrium (\ref{eq:form-V}), the forward and backward reactions occur at equal rates. Similarly, in the present problem, the same configuration must be accessible on both sides of the glass-liquid boundary in order for the process to remain reversible without energy exchange, such as latent heat. Since the liquid phase can access all configurations $(K)$, the glass phase must likewise access the same set of configuration if the process is to be reversible. This can be accomplish by replacing the single cooling process from $T_{g}+$ to $T_{1}$ with a series of infinitesimal cooling and heating steps between the two temperatures. 
After reaching $K_{1}$ at $T_{1}$, the sample is reheated to $T_{g}+$, where the liquid takes generally a different configuration $K_{2}$. Immediately after reaching this configuration, this state of liquid cooled again to $T_{1}$, where the glass takes configuration $K_{2}$. By repeating this cycle of reheating to $T_{g}+$ and cooling back to $T_{1}$ by an arbitrary number $m$ of times, the system successively reaches configurations $K_{3}, K_{4}, \cdots, K_{m}$. In the limit of sufficiently large $m$, in principle, the system explore the entire space $(K)$. 

At this stage, the timescale of equilibrium comes to play a role. Every equilibrium has a finite lifetime given by the relaxation time, Eq.~(\ref{eq:tau}). Complementary to the relaxation time is the growth time ($\tau_{g}$), defined as the time required to reach a new equilibrium \cite{Shirai-SH-Liquids25}. To certificate the time-averaged quantity Eq.~(\ref{eq:t-aveX}) as TC, the sampling time $t_{0}$ must be longer than $\tau_{g}$. Since the glass transition is such a slow process that, within laboratory timescales, it is impossible to explore the whole space of configurations. Consequently, the back-and-forth process is terminated at some finite number $m$. The remaining configurations, from $m$ to $N_{c}$, are unsampled. These unsampled configurations are contained in the integration constant $S_{\rm li}(T_{m}+)$ in Eq.~(\ref{eq:DeltaS-b}), and remains after subtracting by the integration part $\int C \ln T$.
Therefore, the observation of the residual entropy in a given sample does not imply that the sample ``knows" the configurations of other samples. Rather, the residual entropy reflects the outcome of unsampling other configurations during the growth. Indeed, when the glass is formed more quickly in cooling, a larger residual entropy is observed.

\section{Conclusion}
\label{sec:conclusion}
We have established an unambiguous formulation of the third law using Postulation 2. This formulation does not leave exceptions, including glass. It is independent of whether the state is the lowest-energy state or not and does not rely on the traditional distinction between state and metastable states.
To reach this conclusion, the concept of equilibrium was reconsidered. Equilibrium is defined only under a specified set of constraints $\{ \xi_{j} \}$, which restricts the timescale over which the equilibrium maintains. Each constraint specifies a thermodynamic coordinate $X_{j}$, and the set $\{X_{j}\}$ defines the thermodynamic state space ${\mathscr A}$. The entropy value is uniquely determined only on specifying the thermodynamic state space; it must therefore be written as $S^{\mathscr A}$.
From this foundation, the TCs of solids are identified as the time-averaged atomic positions $\bar{\bf R}_{j}$. The set $\{ \bar{\bf R}_{j} \}$ determines the equilibrium configuration of a solid. As temperature approaches zero, only one configuration $\{ \bar{\bf R}_{j}^{(K=1)} \}$ remains thermally accessible within the given constraints. This configuration constitutes the unique equilibrium state at $T=0$. The third law of the zero entropy applies to this state space.

Residual entropy appears when the entropy is evaluated in an extended space including frozen coordinates $\{ \hat{\bf R}_{j}^{(K')} \}$, which were active at high-temperature measurements. The connection these two spaces has been clarified by analyzing the glass transition as an example. Many of longstanding confusions regarding the entropy of glass arise from overlooking the requirement that entropy difference must be evaluated along reversible paths. 
Reversible paths can be found only within the same thermodynamic state space, unless external interventions are introduced. Recognizing this principle resolves the apparent contradiction between zero entropy at absolute zero and the experimentally observed residual entropy.

\section*{Acknowledgments}
The author thanks F.~Belgiorno for valuable discussion on many aspects of the third law, Y.~Oono on the aspect of nonequilibrium states, O. Yamamuro on the experiment aspect of glass, and P. D. Gujrati on the glass transition. 
We also thank Enago (www.enago.jp) for the English language review.
We received financial support from the Research Program of ``Five-star Alliance" in ``NJRC Mater.~\& Dev." (2020G1SA028) from the Ministry of Education, Culture, Sports, Science and Technology of Japan (MEXT).





\bibliography{thermo-refs, glass-refs,info-theory,biology,added}

\begin{thebibliography}{106}
\bibitem [{com({\natexlab{a}})}]{comment1}%
  \BibitemOpen
  \href@noop {} {} ({\natexlab{a}}),\ \bibinfo {note} {today, this view for
  thermodynamics is seriously subjected to criticism. The first law says the
  energy conservation, which holds irrespective of whether the state is
  equilibrium or not. The second law says impossibility of the perpetual-motion
  of engine of the second kind, how the engine utilizes nonequilibrium states.
  These two laws are best applied to engineering applications of energy
  conversion systems, which are far from equilibrium
  \cite{Cengel,Wark}.}\BibitemShut {Stop}%
\bibitem [{Bea({\natexlab{b}})}]{Beauragard-note}%
  \BibitemOpen
  \href@noop {} {} ({\natexlab{b}}),\ \bibinfo {note} {beauragard and Tribus
  wrote: ``There is no way out of the dilemma that equilibrium is defined via
  thermodynamic constructs which constructs were in turn defined for the
  equilibrium state” \cite{Beauragard03}.}\BibitemShut {Stop}%
\bibitem [{com({\natexlab{d}})}]{comment-few-GB}%
  \BibitemOpen
  \href@noop {} {} ({\natexlab{d}}),\ \bibinfo {note} {gyftopoulos and Berreta
  state in their book \cite{Gyftopoulos} (p.~119) that the state principle (it
  can be read as the notion of state variable in the present context) is
  derived as a rigorous consequence of the first and second laws, not as a
  consequence either of difficulties related to exact calculations and lack of
  knowledge, or a need to describe complicated physical problems by a few gross
  macroscopic averages.}\BibitemShut {Stop}%
\bibitem [{com({\natexlab{e}})}]{comment5}%
  \BibitemOpen
  \href@noop {} {} ({\natexlab{e}}),\ \bibinfo {note} {cooling causes an
  increase in the entropy of the environment due to the emission of heat from a
  refrigerator. But this was already compensated by the decrease in $\int C
  d\ln T$ of glass. This is the reversible part. Further decrease in $S$
  occurs, which is the issue to be answered.}\BibitemShut {Stop}%
\end{thebibliography}



\end{document}